\documentstyle[12pt]{article}
\begin{document}

\title{On the Faddeev-Popov determinant in Regge calculus}
\author{V.M.Khatsymovsky \\
 {\em Budker Institute of Nuclear Physics} \\ {\em
 Novosibirsk,
 630090,
 Russia} \\ {\em E-mail address: khatsym@inp.nsk.su}}
\date{}
\maketitle
\begin{abstract}
The functional integral measure in the 4D Regge calculus
normalised w.r.t. the DeWitt supermetric on the space of
metrics is considered. The Faddeev-Popov factor in the
measure is shown according to the previous author's work
on the continuous fields in Regge calculus to be generally
ill-defined due to the conical singularities. Possible
resolution of this problem is discretisation of the gravity
ghost (gauge) field by, e.g., confining ourselves to the
affine transformations of the affine frames in the
simplices. This results in the singularity of the
functional measure in the vicinity of the flat background,
where part of the physical degrees of freedom connected
with linklengths become gauge ones.
\end{abstract}

\newpage

The functional integral approach to quantisation of Regge
calculus based on diffeomorphisms has been started probably
in ref. \cite{JN} and further developed by Menotti and
Peirano in a number of papers, refs. \cite{MP1,MP2}.
One of basic ingredients of this approach is Faddeev-Popov
determinant which represents the gauge volume of geometries
over which functional integration is performed. It is
determinant of some second order differential operator
acting on the space tangential to the diffeomorphism group,
and the elements of this space, vectors of infinitesimal
frame deformations, thus play the role of ghost fields.

Logarithm of the Faddeev-Popov factor gives, in fact,
effective action of the ghost field. This field is
continuous
one. Meanwhile, it was shown in the recent author's paper
ref. \cite{Kha} that such action can suffer from the
infinities unremovable by the UV renormalisation procedure.
This can be checked by studying the conformal (trace)
anomaly of this field which due to occurence of curvature
bilinears and $\delta$ -function-like nature of Regge
curvature turns out to contain such terms.

In the present letter we calculate trace anomaly of
the (vector) Faddeev-Popov ghost field for the particular
choice of free constant in the DeWitt metric, namely that
used in ref. \cite{JN}. The divergent part turns out to be
nonzero and, moreover, of the same sign as the divergent
parts in the case of matter fields studied in ref.
\cite{Kha}, therefore one hardly can hope that any
cancellation occurs. Of course, in the case of conformally
noninvariant fields such as vector ghost one, there are
also contributions to the trace of stress-energy tensor
depending on the state of the field, therefore our
argumentation here is that the cancellation of
singularities can occur only for special choice of the
state of the field which is equivalent to imposing a new
constraint not present in the underlying continuous theory.

Aiming at constructing finite anzats for the measure we
propose to confine ourselves by a discrete number of gauge
degrees of freedom corresponding to the affine
transformations of the affine frames inside the simplices.
If we adopt this, the measure is singular in the vicinity
of the flat background. This is because gauge degrees of
freedom are absorbed in this background by the physical
ones (linklength variations) and Gaussian normalisation
integral for the measure diverges.

Let us begin with some definitions of refs.
\cite{JN,MP1,MP2} of interest. The DeWitt supermetric on
the infinitesimal deformations of metric $\delta
g_{ik}$ is chosen in the form \cite{JN}
\begin{equation}
\label{DeWitt}
(\delta g,\delta g)=\int dx G^{iklm}\delta g_{ik}
\delta g_{lm}
\end{equation}
with
\begin{equation}
\label{G}
G^{iklm}={1\over 2}\sqrt{g}(g^{il}g^{km}+g^{im}g^{kl}-
g^{ik}g^{lm}).
\end{equation}
For the infinitesimal coordinate shifts $\xi_\mu$ (ghost
field) and gauge metric deformations $\delta g_{ik}=\nabla_
i\xi_k+\nabla_k\xi_i$ Gaussian integral normalisation of
the measure leads, in particular, to computation of the
determinant of some second order differential operator so
that we need to compute effective action for the field
$\xi$ with Lagrangian density proportional to
\begin{equation}
\label{A}
\xi_i (\delta_k^i \nabla^2+R^i_k)\xi^k
\end{equation}
This operator differs by the sign at $R^i_k$ from the
analogous one in the electromagnetism, and we calculate
the trace anomaly for it using asymptotic expansion of
$<x|\exp{(tA)}|x>$ \cite{Sch,CD,Chr} where operator $A$ just
appears in (\ref{A}). The trace anomaly of interest is
given by $t^{-n}$ term at $n$ = 0. The asymptotic expansion
itself can be found representing $A$ as the sum of
electromagnetic operator $B$ (differing from $A$ by the sign
of $R^i_k$) and $2{\cal R}$, the ${\cal R}$ denoting
matrix $R^i_k$, and by disentangling exponents. The
explicit expressions look like
\begin{equation}
\label{exp-t-A}
\exp{(tA)}=\{1+2{\cal R}t+t^2(2{\cal R}^2+[B,{\cal R}])
+O(t^3)\}\exp{(tB)}
\end{equation}
and
\begin{eqnarray}
<x|\exp{(tB)}|x>^i_k & = & {1\over (4\pi t)^2}\left\{
\delta ^i_k+
t\left ({1\over 6}R\delta^i_k-R^i_k\right )+t^2\left [
\delta^i_k\left ({1\over 72}R^2\right.\right.\right.
\nonumber \\
 & - & \left.\left.\left. {1\over 180}
R_{lm}R^{lm}+{1\over 180}R_{lmpq}R^{lmpq}-{1\over 30}\Delta
R\right )+{1\over 15}R^l_mR^{im}_{kl}\right.\right.\nonumber\\
\label{exp-t-B}
& + & \left.\left.{13\over 30}R_{lk}R^{li}-{1\over 6}RR^i_k
-{1\over 12}R^{mp}_{kl}R^{il}_{mp}+{1\over 6}\Delta R^i_k
\right ]\right\}+...
\end{eqnarray}
One should be careful when calculating trace of the product
of the
differential operators and $\exp{(tB)}$ in the RHS of
(\ref{exp-t-A}) and use for this purpose the
$<y|\exp{(tB)}|x>$ differing from (\ref{exp-t-B}) by
appearance, in the required orders, of the heat kernel
exponent
$$\exp{\left [-{(y-x)^2\over 4t}\right ]}$$
in the RHS of (\ref{exp-t-B}). It is useful to write out
the result for the trace anomaly for the operator $B+
\epsilon {\cal R}$,
\begin{eqnarray}
\label{T-mu-mu}
<T^\mu_\mu> & = & {1\over (4\pi)^2}\left [{1\over 12}
\left (\epsilon -{2\over 3}\right )R^2+{1\over 12}
\left (\epsilon +{1\over 5}\right )\Delta R
+\left ({1\over 4}\epsilon^2-{1\over 2}\epsilon
+{43\over 180}\right )R_{ik}R^{ik}\right.\nonumber\\
& - & \left.{11\over 360}R_{iklm}R^{iklm}\right ].
\end{eqnarray}
At $\epsilon$ = 0 we reproduce the trace anomaly for the
electromagnetic operator (not for the electromagnetic field
itself: for this field one should also subtract scalar ghost
contribution); at $\epsilon$ = 2 we have this for the vector
ghost operator of interest.

According to the previous author's paper \cite{Kha} the
curvature bilinears are ill-defined on the 4D Regge lattice
and have unremovable by the UV renormalisation infinite
parts;
substituting the latter parts into the $<T^\mu_\mu>$ we
find that the whole singular part is proportional to
\begin{equation}
{1\over 4}\epsilon^2-{1\over 3}\epsilon+{1\over 15}.
\end{equation}
This is 1/15 and 2/5 for the electromagnetic and the
considered ghost operator, respectively; in this scale
the electromagnetic field itself gives 1/30, i.e. of the
same sign. As it was found in ref. \cite{Kha}, contributions
from the other matter fields have the same sign. This means
that it is not possible to have situation with the mutual
cancellation of the ill-defined contributions from the
different fields.

The singularity of the Faddeev-Popov factor in the measure
makes us to reduce the continuum number of the gauge
degrees of freedom to the discrete number; in fact,
the only natural choice on the Regge lattice is to
introduce gauge field as that describing infinitesimal
changes $\xi_a^i$ of the coordinates $x_a^i$ of the
vertices $a$. The coordinate frame is piecewise-affine and
uniquely defined inside each 4-simplex $\sigma$ by the
coordinates of it's vertices. We are interested in the
change of metric $\delta_{ik}$ due to variations of both
vertex coordinates and the lengths squared $s_{(ab)}$ of the
links connecting vertices $a$ and $b$. The supermetric form
(\ref{DeWitt}) on this change can be most conveniently
written using the so-called edge components $\delta g_{(ab)}$
of the symmetric second rank tensor \cite{PW}, here
$\delta g_{ik}$,
\begin{equation}
\delta g_{(ab)}\equiv l^i_{ab}l^k_{ab}\delta g_{ik}
\end{equation}
where $l^i_{ab}\equiv x^i_a-x^i_b$. In terms of length
and coordinate variations these can be found in a simplex
$\sigma$ to be
\begin{equation}
\label{edgecomp}
\delta g_{\sigma (ab)}=\delta s_{(ab)}-2g_{\sigma ik}l^i_{ab}
(\xi^k_a-\xi^k_b).
\end{equation}
Here $g_{\sigma ik}$ is the metric inside $\sigma$. With
these notations according to ref. \cite{PW} we can find
\begin{equation}
(\delta g,\delta g)=\sum_\sigma\sum_{(ab),(cd)\subset\sigma}
\left [{1\over V^4_\sigma}{\partial V^2_\sigma\over
\partial s_{(ab)}}{\partial V^2_\sigma\over
\partial s_{(cd)}}-{2\over V^2_\sigma}{\partial^2 V^2_\sigma\over
\partial s_{(ab)}\partial s_{(cd)}}\right ]\delta g_{\sigma (ab)}
\delta g_{\sigma (cd)},
\end{equation}
$V_\sigma$ being the volume of the simplex. When
normalising the measure w.r.t. the DeWitt metric we perform
(Gaussian) integration of $\exp{(-(\delta g,\delta g))}$
over $d\xi$, $d\delta s$.

If Regge spacetime is flat, the metric
$g_{\sigma ik}$ entering (\ref{edgecomp}) can be chosen the
same for all the simplices $\sigma$. Therefore $\xi^k_a-
\xi^k_b$ can be absorbed by $\delta s_{(ab)}$. This means
that (generally convergent) Gaussian integral becomes
divergent since the form $(\delta g,\delta g)$ becomes
degenerate.

Singularity of the measure suggested in the completely 
discrete Regge spacetime in the vicinity of the
flat background has a parallel with the author's paper
\cite{Kha1} where the continuous time (3+1)D Regge calculus
has been formulated in the canonical form. In the latter
case the functional integral constructed according to the
Dirac prescription turns out to be singular in the vicinity
of the flat background too, and the reason for that is
connected with converting a part of the second class
constraints into the first class ones, i. e. again has the
symmetry origin.

The possibility which still remains to be studied is using
the form of the DeWitt metric more general than (\ref{G})
where the coefficient at the term $g^{ik}g^{lm}$ is arbitrary
constant as in \cite{MP1,MP2}. The possibility a'priory
exists that this coefficient can be chosen to make singular
part in the trace anomaly vanishing. Of course, this does
not necessarily mean that the effective action is
well-defined, but admit such possibility. In this case the
operator $A$ is the sum of not only Laplacian and curvature
term, but also of the term $\sim\nabla^i\nabla_k$. This
term makes disentangling the exponent like (\ref{exp-t-A})
much more difficult (infinite number of terms is required
there) and, probably, expansion around the electromagnetic
case does not give any advantage. Recently corresponding
calculations are in progress.

\bigskip

This work was supported in part by the RFBR grant
No. 96-15-96317.

\end{document}